\documentclass[]{interact}

\usepackage{epstopdf}% To incorporate .eps illustrations using PDFLaTeX, etc.
\usepackage[caption=false]{subfig}% Support for small, `sub' figures and tables

\usepackage{braket}
\usepackage[numbers,sort&compress]{natbib}
\usepackage{url}% Citation support using natbib.sty
\bibpunct[, ]{[}{]}{,}{n}{,}{,}% Citation support using natbib.sty
\renewcommand\bibfont{\fontsize{10}{12}\selectfont}% Bibliography support using natbib.sty
\usepackage{xspace}
\usepackage{color}

\newcommand{\HPhi}{$\mathcal{H}\Phi$\xspace}
\newcommand{\Komega}{K$\omega$\xspace}

\makeatletter% @ becomes a letter
\def\NAT@def@citea{\def\@citea{\NAT@separator}}% Suppress spaces between citations using natbib.sty
\makeatother% @ becomes a symbol again

\theoremstyle{plain}% Theorem-like structures provided by amsthm.sty

\theoremstyle{definition}

\theoremstyle{remark}

\begin{document}

%\articletype{ARTICLE TEMPLATE}% Specify the article type or omit as appropriate

\title{Project For Advancement of Software Usability in Materials Science}

\author{
\name{Kazuyoshi Yoshimi\textsuperscript{a}\thanks{CONTACT Kazuyoshi Yoshimi. Email: k-yoshimi@issp.u-tokyo.ac.jp}, Yuichi Motoyama\textsuperscript{a}, Tatsumi Aoyama\textsuperscript{a}, Mitsuaki Kawamura\textsuperscript{b} and Naoki Kawashima\textsuperscript{a}}
\affil{\textsuperscript{a}The Institute for Solid State Physics, The University of Tokyo, Chiba 277-8581, Japan; \textsuperscript{b}Faculty of Engineering, Yokohama National University, Yokohama 240-8501, Japan}
}

\maketitle

\begin{abstract}
The Institute for Solid State Physics (ISSP) at The University of Tokyo has been carrying out a software development project named ``the Project for Advancement of Software Usability in Materials Science (PASUMS)". 
Since the launch of PASUMS, various open-source software programs have been developed/advanced, including ab initio calculations, effective model solvers, and software for machine learning.
We also focus on activities that make the software easier to use, such as developing comprehensive computing tools that enable efficient use of supercomputers and interoperability between different software programs.
We hope to contribute broadly to developing the computational materials science community through these activities.
\end{abstract}


\section{Introduction}

In recent years, topics such as machine learning and quantum computing have become widely recognized in society for the importance of computation in a broad sense.
Naturally, these trends have also significantly impacted the field of condensed matter science.
Innovative computational methods in condensed matter science are invariably based on a deep understanding of physical systems, and the essential relationship between computation and physics is obvious to researchers, and this relationship has become more explicitly acknowledged. 
This paper introduces the efforts of The Institute for Solid State Physics (ISSP) at The University of Tokyo in computational materials science, focusing on a project for supporting the computational materials science community through the development/improvement of software.

One of the key initiatives supporting the computational materials science community at ISSP is the nationwide joint use of supercomputer hardware \cite{ActivityReport}. 
On the other hand, with the advancement of computer architectures, the human cost of developing original code has increased. The Project for Advancement of Software Usability in Materials Science (PASUMS)~\cite{pasums} was launched in 2015 to address this issue. 
This initiative was inevitable, as many ISSP supercomputer users use codes developed independently.
In addition, the highly efficient computational algorithms and the programs that implement them reflect the characteristics of the physical systems under study, and making them publicly accessible is an efficient way of disseminating the essential ideas of condensed matter theories to the world.
In this paper, we would like to provide an overview of the PASUMS project and introduce the software developed through the project with a few applications.

\section{Overview of PASUMS}

PASUMS is a program conducted by ISSP to develop and enhance software that is important in condensed matter physics and materials science, and is expected to be used on ISSP supercomputer systems.
It is part of the nationwide joint-use program of the supercomputer, which provides computational resources to domestic researchers, supports software development, and deals with the increasing complexity of modern large-scale parallel computer systems.
Software to be developed is called for annually. The steering committee of the joint use program examines proposals, and two or three are accepted every fiscal year.
In the PASUMS program, we develop and enhance software functionality, improve user interfaces, and prepare documentation and tutorials.
We also support disseminating the software, installing it on the ISSP supercomputer systems, holding hands-on lectures, building dedicated websites, and writing software papers.
The program aims to contribute to the advancement of the field by sharing the products as community codes.
The deliverables are distributed as open-source software, and users can use them on their sites and extend them to solve their target problems.
We believe it essential that the derivative works remain community codes.
For this purpose, we adopt the copyleft license for the deliverable software. The software packages developed in the PASUMS program are, in principle, distributed under the GNU Public License (GPL)~\cite{gpl}, except for the library packages, in which case the Lesser GPL~\cite{lgpl} or Mozilla Public License (MPL)~\cite{mpl} are adopted.

By the end of the school year 2024, 21 projects have been adopted. Table~\ref{table-soft-list} lists the software that PASUMS has handled so far. Most of the applications are intended for computation related to some specific type of physical systems: four related to the first-principles calculations, six to the quantum lattice model solvers, and five to other areas such as machine learning. In the next section, we will pick up some applications among them.
There is another category of applications intended to assist the usage of other applications, though not listed in Table~\ref{table-soft-list}.
For example, we enhanced a software package, StdFace~\cite{stdface}, that generates model definition files for the quantum lattice model solvers \HPhi~\cite{HPhi}, mVMC~\cite{mVMC}, and H-wave~\cite{Hwave} from a simple description so that it accepts the output of the software package RESPACK~\cite{RESPACK} that derives effective models.
It enables seamless processes from the first-principles calculations through deriving the low-energy effective Hamiltonians to analyzing the resulting effective models.
Yet another example of assisting applications is the software package, moller~\cite{htp_tools}, which generates job scripts for batch job schedulers that integrate large-scale computational resources to perform exhaustive calculations.
It allows for the swift construction of databases for highly accurate models to estimate physical quantities using data science techniques.
We believe these efforts should contribute to accelerating the discovery of new functional materials through materials informatics.

\section{Developed/Enhanced Software}

\subsection{First-Principles Calculation Related}

First-principles calculations, which simulate molecules, solids, surfaces, interfaces, and nanostructures have become a valuable research tool for condensed matter theorists and experimental researchers. These calculations serve as standalone analysis tools and form the basis for combined analyses with classical molecular dynamics, which handle larger scales, and quantum lattice models, which incorporate high-precision correlation effects (discussed in the next section). As a result, they account for a significant portion of the use of the ISSP supercomputer. Programs for performing first-principles calculations, such as VASP~\cite{vasp}, Quantum ESPRESSO~\cite{QUANTUMESPRESSO,Giannozzi2009,Giannozzi2017}, and OpenMX~\cite{OpenMX}, are packaged and maintained to be user-friendly for researchers. Below, we describe the additions and improvements to first-principles calculation packages through PASUMS.

\subsubsection{OpenMX (2015)}
\begin{figure}[!h]
    \centering
    \includegraphics[width=.45\textwidth]{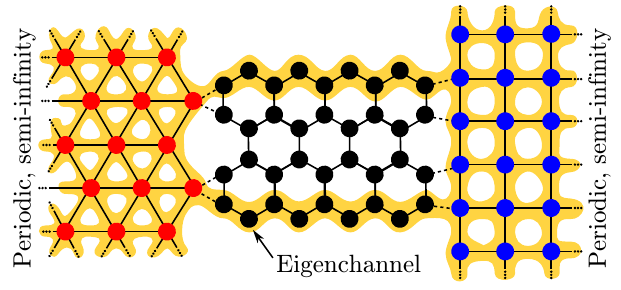}
    \caption{
      Schematic of electrical conductance and eigenchannels (yellow) in a nanostructure. Periodic boundary conditions and Bloch's theorem are applied in the vertical and front-back directions. In contrast, the periodic structure of each electrode material continues semi-infinitely in the left-right direction.
    }
    \label{fig_channel}
\end{figure}

OpenMX (Open source package for Material eXplorer) is a first-principles calculation package using numerical localized basis sets~\cite{Ozaki2003,Ozaki2004,Ozaki2005}. It supports standard functions such as band calculations, structural optimization, and Born-Oppenheimer molecular dynamics calculations based on Density Functional Theory (DFT), in addition to large-scale calculations with computational costs proportional to the number of atoms (as opposed to the cubic scaling in typical DFT)~\cite{PhysRevB.74.245101}. It also supports electrical conductance calculations for nanoscale junctions using the non-equilibrium Green's function method~\cite{PhysRevB.81.035116}. An eigenchannel analysis~\cite{PhysRevB.76.115117}, which aids in physically interpreting the conductance and current obtained in these electrical conductance simulations, is also available. Eigenchannels are a superposition of wavefunctions in nanoscale junctions obtained by diagonalizing the transmission matrix, yielding eigenconductance. One can investigate the paths electrons take by examining the real-space distribution of channels with large eigenconductance (Figure \ref{fig_channel}).

\subsubsection{ESM-RISM (2021)}
\begin{figure}[!tb]
    \centering
    \includegraphics[width=.45\textwidth]{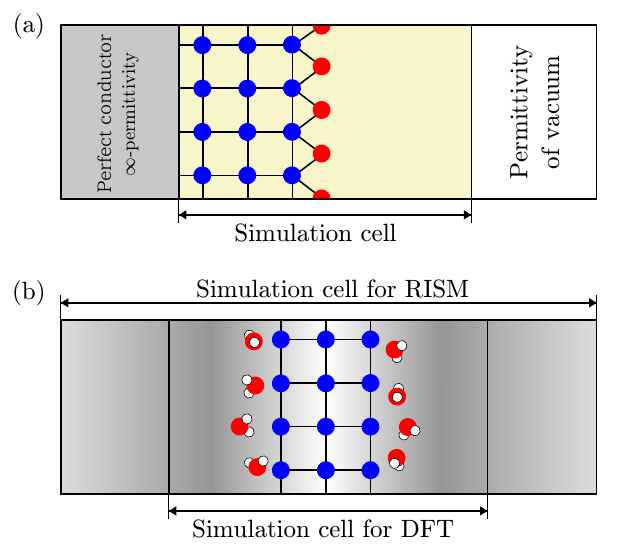}
    \caption{
      (a) Schematic of the ESM method. The Poisson equation can be solved analytically in regions of perfect conductors and vacuum, and these results are connected to the electrostatic potential within the simulation cell for efficient non-periodic system calculations. The Kohn-Sham equation in each self-consistent step is solved within the simulation cell under conventional periodic boundary conditions. (b) Schematic of the ESM-RISM method. The solvent density is represented in grayscale. The Kohn-Sham and RISM equations are solved in separate simulation cells, interconnected through the electrostatic potential.
    }
    \label{fig_esm}
\end{figure}

The ESM-RISM software module~\cite{ESMRISM} is an extension implemented within Quantum ESPRESSO~\cite{QUANTUMESPRESSO}, a widely used open-source software package for first-principles electronic structure calculations. It is designed to simulate systems exhibiting broken periodicity in one dimension, such as solid-liquid interfaces encountered in battery electrodes and dielectric surfaces. Traditional approaches typically use slab models with periodic vacuum regions, which can lead to non-physical interactions due to periodic boundary conditions.
To overcome this issue, the Effective Screening Medium (ESM) method~\cite{PhysRevB.73.115407} was proposed, where the Kohn-Sham equation for electrons is solved under periodic boundary conditions, while the Poisson equation for the electrostatic field is solved under open boundary conditions, representing semi-infinite vacuum or perfect conductor conditions.
The ESM-RISM method further enhances the ESM approach by incorporating the Reference Interaction Site Model (RISM), a classical liquid theory, enabling efficient and accurate simulations of electrolyte distributions and electric double layers at electrode-electrolyte interfaces under bias voltage conditions (Figure \ref{fig_esm}(a)). This capability is particularly beneficial for modeling electrochemical systems such as fuel cell electrodes~\cite{PhysRevB.96.115429}.

Under the PASUMS, the ESM-RISM module was significantly improved by introducing flexible spatial meshing schemes. Initially, the ESM-RISM method employed a common spatial mesh for electronic (DFT) and classical (RISM) equations, suitable primarily for thin electric double layers. PASUMS introduced independent spatial discretizations for the DFT and RISM equations using Fourier interpolation methods. This allowed for distinct mesh densities and cell sizes tailored to each equation (Figure \ref{fig_esm}(b)), significantly broadening the method's application scope. Comprehensive manuals and tutorials were also provided to promote the practical adoption and effective use of this improved ESM-RISM implementation.

\subsubsection{RESPACK (2018)} 

\begin{figure}[!tb]
  \centering
  \includegraphics[width=.45\textwidth]{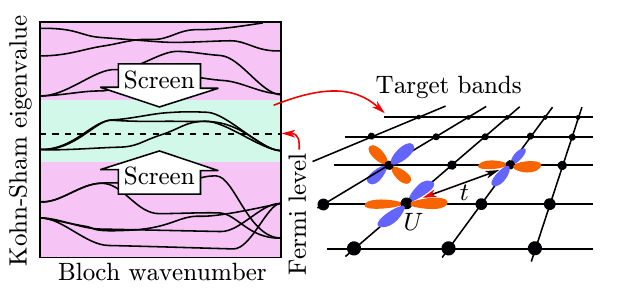}
  \caption{
    Schematic of downfolding. From the band structure and Kohn-Sham orbitals obtained by first-principles calculations, parameters such as hopping integrals \( t \) and Coulomb integrals \( U \) of the Hubbard model are calculated by focusing only on the states near the Fermi surface (target bands) using maximally localized Wannier functions and constrained random phase approximation. Contributions from orbitals other than the target bands are included as screening to the atomic potential and electron-electron Coulomb interactions.
  }
  \label{fig_downfold}
\end{figure}

While first-principles calculations address various problems, systems with strong electron correlations, such as high-temperature superconductors made of copper oxides, require theoretical frameworks beyond standard first-principles methods. One approach is to construct lattice models, such as the Hubbard model, from first-principles calculation results (downfolding, see Figure \ref{fig_downfold}), and apply high-precision methods, such as exact diagonalization and quantum Monte Carlo methods, to these models. The crucial point is not losing the physical essence of the modeling process. RESPACK~\cite{RESPACK-paper} is a program package designed to generate such models, ensuring these properties using maximally localized Wannier functions~\cite{RevModPhys.84.1419} and constrained random phase approximation~\cite{PhysRevB.70.195104}. PASUMS has developed interfaces to integrate with several high-precision lattice model solvers seamlessly.

\subsection{Quantum Lattice Model Solver Related}

Various quantum lattice models, such as the Hubbard model, where electrons hop between lattice sites while interacting with each other, and the Heisenberg model, where spins fixed at lattice sites interact, have been proposed and studied. These models' ground and low-lying excited states are the main research targets, but obtaining exact solutions is impossible for most cases, except for a few exceptions. Therefore, various numerical methods have been proposed~\cite{StronglyCorrelatedSystemsNumericalMethods}, and corresponding software has been developed. Here, we briefly introduce some representative methods and the software developed or enhanced through PASUMS.

\subsubsection{H-wave (2022)}
Mean-field approximations generally reduce the original many-body problem to a single-body problem that deals with the fluctuations of physical quantities up to first order.
Although systematically controlling the degree of approximation is difficult, the mean-field approximations are widely applicable and computationally inexpensive compared to other methods, making them useful for intuitive understanding, such as examining rough phase diagrams over a wide range of parameters, or applying for screening before precise calculations.

H-wave~\cite{Hwave} is a software package that implements the unrestricted Hartree-Fock (UHF) method for quantum lattice models, and the random phase approximation, allowing for the calculation of susceptibilities for single-body physical quantities.
It supports the UHF method both in the real space and the wavenumber space, which is expected to lead to a significant speed-up for systems with short-range order parameters.
It also supports finite temperature calculations.

\begin{figure}[!tb]
  \centering
  \includegraphics[width=.7\textwidth]{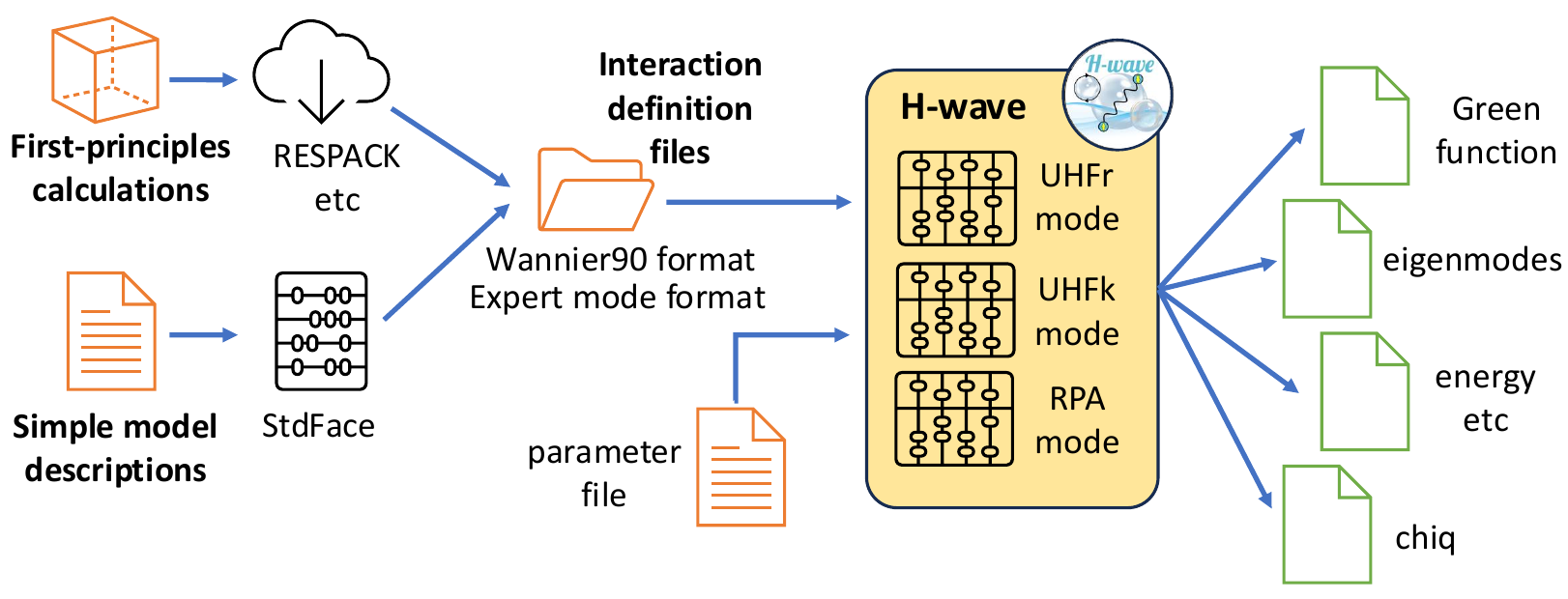}
  \caption{
    Schematic flow of calculations using H-wave.
    The users prepare the interaction definition files in the Wannier90 format or the expert-mode format and the input parameter files.
    The interaction definition files can be generated from a simple description by StdFace, or derived from the first-principles calculations.
    The results are stored in the output files according to the input parameters, including the expectation values of the physical observables, the Green's functions, and other data for further analyses.
  }
  \label{fig:hwave}
\end{figure}

The input files describing the one-body and two-body interactions are based on the Wannier90 format.
This allows for a smooth connection with the software packages that derive the effective models from the first-principles calculations, such as RESPACK, to analyze them with H-wave. (Figure~\ref{fig:hwave})

The following functionalities and features are suggested as future extensions:
(a) calculations of quantities corresponding to dynamic susceptibility measured in experiments,
(b) evaluation of the instability of superconducting transition by solving the linear Eliashberg equation considering charge and spin fluctuations as pairing interactions, and
(c) more samples such as the calculation with spin-orbit interactions.

\subsubsection{DCore (2017, 2024)}

\begin{figure}
\includegraphics[width=\linewidth]{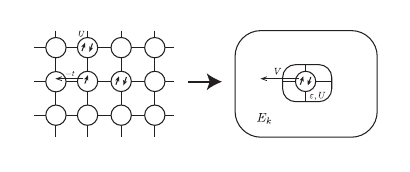}
\caption{
Schematic figure of the map from a lattice Hamiltonian to an effective impurity model in the DMFT.
}
\label{fig:DMFT}
\end{figure}

Dynamical mean-field theory (DMFT)~\cite{Georges1996, Kotliar2006} is one of the most powerful tools for investigating strongly correlated electron systems.
The DMFT maps a lattice model to an impurity model in an effective medium (Figure~\ref{fig:DMFT}) and solves this model self-consistently as the mean-field approximation.
This method treats the imaginary-time Green's function and self-energy as the main physical quantities, and it can calculate the dynamical properties of the system like one-particle spectral functions, which are directly compared with the experimental results by angle-resolved photoemission spectroscopy (ARPES).
In addition to the theoretical lattice model such as the Hubbard model, the DMFT can be applied to real materials by combining with DFT calculations.
This approach, called DFT+DMFT, has been widely used to investigate the electronic structure of strongly correlated materials such as transition-metal oxides~\cite{Kotliar2006}.

DCore (abbreviation of the ``integrated DMFT software for CORrelated Electrons'')~\cite{DCore} was released as DMFT software in 2017 with the help of PASUMS.
DCore implements the main DMFT loop and leaves the impurity solver to an external program, which enables the use of various impurity solvers.
In the latest version (4.1.0), DCore supports four algorithms for the impurity solver: The continuous-time quantum Monte Carlo (CTQMC) method, the exact diagonalization, the Hubbard-I approximation, and the non-interacting limit (zero self-energy limit).
For the CTQMC method, three programs are supported: TRIQS/cthyb~\cite{TRIQS-CTHYB-paper, TRIQS-CTHYB-web}, ALPS/CT-HYB~\cite{ALPS-CTHYB-paper, ALPS-CTHYB-github}, and ALPS/CT-HYB-SEGMENT~\cite{ALPS-CTHYB-SEG-paper, ALPS-CTHYB-SEG-github}.
For the exact diagonalization, pomerol~\cite{pomerol-zenodo} is supported.
For the Hubbard-I approximation, pomerol and TRIQS/hubbard-I~\cite{TRIQS-HubbardI-web} are supported.
By installing these external programs, users can choose the most suitable impurity solver for their target systems.

Since its release, several features have been added to DCore, such as more solvers and postprocesses.
In the fiscal year 2024, PASUMS made the postprocesses of DCore more organized.
One of the postprocesses is Bethe-Salpeter equation (BSE) solver~\cite{Tagliavini2018}, which calculates the two-body susceptibility
\begin{equation}
\chi_{ij,kl}(\boldsymbol{q},\Omega_m) = \frac{1}{N}\sum_{\boldsymbol{r}} \int_0^\beta d\tau \Braket{c^\dagger_i(\boldsymbol{r},\tau)c_j(\boldsymbol{r},\tau)c^\dagger_k(\boldsymbol{0},0)c_l(\boldsymbol{0},0)}e^{i\Omega_m\tau}e^{-i\boldsymbol{q}\cdot\boldsymbol{r}},
\end{equation}
where $\boldsymbol{r}$ denotes the position of the unitcell and $i,j,k,l$ are the combined spin and orbital indices in the unitcell.
From the DMFT+BSE calculation, for example, the location of the critical point can be estimated as the zero point of the inversed susceptibility $\chi^{-1}$.

\subsubsection{\HPhi (2015, 2016, 2017) and \Komega (2016)}
For quantitative comparisons with experimental data, numerical exact diagonalization (ED) of quantum lattice systems is one of the most reliable methods for small systems, as it avoids approximations. This approach enables detailed analysis of quantum systems and serves as a benchmark for other numerical techniques. 
With the increasing availability of parallel computing infrastructures featuring distributed-memory architectures and narrow bandwidths, there is a growing demand for efficient, user-friendly, and highly parallelized diagonalization software. 

To address this need, \HPhi~\cite{HPhi-paper, HPhi-paper-updated} was developed.
\HPhi yields eigenstates of any given lattice fermion Hamiltonian defined on a finite set of lattice points that consists of hopping terms and multi-body interactions. As a special case, it can also be applied to spin systems. The typical supported models are listed as follows:
\begin{itemize} 
\item Hubbard and Heisenberg models, 
\item Multi-band extensions of the Hubbard model, 
\item Models with SU(2)-symmetry-breaking exchange interactions, such as Dzyaloshinskii-Moriya and Kitaev interactions, 
\item Kondo lattice models that describe itinerant electrons coupled with quantum spins. 
\end{itemize}
\HPhi enables the computation of numerous physical quantities, including: 
\begin{itemize} 
\item Internal energy at zero and finite temperatures, 
\item Temperature-dependent specific heat, 
\item Charge and spin structure factors, 
\item Optical spectra and other dynamical properties. 
\end{itemize}
These features make \HPhi a versatile and powerful tool for researchers across various fields, including experimentalists who seek to validate their findings with theoretical models.

The development of \HPhi spans multiple phases. \HPhi ver. 1 was completed as part of the fiscal year 2015 project, while \HPhi ver. 2 was developed during the fiscal year 2016. \HPhi ver. 2 integrates seamlessly with the numerical library \Komega~\cite{Komega-paper}, which was developed in parallel under the same project. \Komega provides advanced numerical routines for linear algebra, significantly enhancing the computational efficiency of \HPhi. For example, the shifted bi-conjugate gradient (sBiCG) method was implemented in \HPhi using \Komega, enabling efficient computation of dynamical Green's functions and excitation spectra. Furthermore, the locally optimal block preconditioned conjugate gradient (LOBPCG) method was incorporated, allowing for the simultaneous computation of multiple low-energy eigenvalues and eigenvectors in a single calculation.
In the fiscal year 2017, \HPhi was extended to include real-time evolution capabilities. This feature allows researchers to study non-equilibrium dynamics in quantum many-body systems, a rapidly growing area of interest in quantum technologies and experiments. These advancements position \HPhi as a comprehensive tool for investigating quantum systems' equilibrium and non-equilibrium properties.

\subsubsection{DSQSS (2018)}
\begin{figure}
\includegraphics[width=\linewidth]{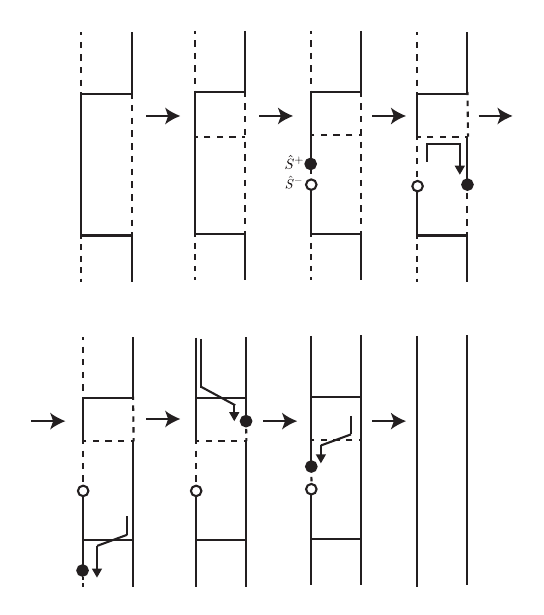}
\caption{
Illustration of the directed loop algorithm for $S=1/2$ spin model.
Vertical solid lines and dashed lines denote upspin and downspin, respectively.
First, vertices (dashed horizontal lines) are generated.
Next, a pair of $\hat{S}^+$ (black circle) and $\hat{S}^-$ (white circle) operators are inserted.
Then, $\hat{S}^+$ moves along lines while flipping spins,
and returns where $\hat{S}^-$ is and removed.
}
\label{fig:WL}
\end{figure}

Path integral Monte Carlo (PIMC) method~\cite{Gubernatis2016} maps the partition function $Z=\mathrm{Tr}e^{-\beta H}$ of a $D$-dimensional quantum system onto that of a ($D+1$)-dimensional classical system using path integrals, and samples paths (``world-lines'') with their weight.
This powerful method allows us to calculate the thermal expectation values of physical quantity exactly within statistical error regardless of the spatial dimension and/or the system size as long as the system is free from the infamous sign problem.
Near the critical point, the convergence of the PIMC method with local update of the world lines becomes slow (critical slowing down).
For discrete space systems, several global update (cluster update) methods have been developed to solve the critical slowing down, for example, the directed loop algorithm~\cite{Syljusen2002}.
Figure~\ref{fig:WL} shows how the directed loop algorithm updates the world-line configuration globally.
Therefore, the PIMC methods with cluster updates have been widely used for studying the critical phenomena of quantum lattice problems.

DSQSS~\cite{DSQSS} is software implementing path integral Monte Carlo methods for quantum lattice systems.
DSQSS can deal with symmetry-broken systems such as spins under magnetic field well because it implements the directed loop algorithm and the parallel worm algorithm~\cite{MasakiKato2014}.
In 2018, PASUMS enhanced the user experience of DSQSS by improving the installation process and developing utility tools for easily generating input files.
By using the input generator tools, the input files describing lattice models,
\begin{itemize}
    \item general spin XXZ model
    \item hardcore and softcore Bose-Hubbard model
\end{itemize}
on
\begin{itemize}
    \item hypercubic lattice in any spatial dimension
    \item triangular lattice
    \item honeycomb lattice
\end{itemize}
are generated.
DSQSS can calculate the thermal average (canonical average) of several observables such as
\begin{itemize}
\item energy and specific heat,
\item magnetization (particle density) and susceptibility,
\item spin-spin (density-density) correlation in space-time.
\end{itemize}

\subsubsection{mVMC (2016)}
The variational method based on Ritz's variational principle searches for the ground state $\Psi(\theta^*)$ by finding the parameters $\theta^*$ that minimizes the energy $E(\theta)$ of a parameterized (many-body) trial wavefunction $\Psi(\theta)$.
The variational Monte Carlo (VMC) method evaluates the average of an arbitrary operator $\hat{A}$ for the given wave function,
$\braket{A}_\theta = \braket{\Psi(\theta)|A|\Psi(\theta)} \Big/ \braket{\Psi(\theta)|\Psi(\theta)}$,
by the Markov chain Monte Carlo method.
Note that the VMC method does not suffer from the sign problem because it does not sample the wave function directly.
In the variational methods,
the accuracy and computational cost of the approximation can be controlled
by varying the parameter set and the numbers that construct the trial wavefunction.
For example, the multi-variable variational Monte Carlo method~\cite{Tahara2008} is based on the trial wavefunction using a combination of a one-body wavefunction written as a Pfaffian-Slater determinant and a few additional factors, such as the Jastrow and Gutzwiller factors, to better represent electronic correlations.

mVMC~\cite{mVMC} is software implementing the multi-variable variational Monte Carlo method for strongly correlated electron systems.
mVMC takes similar input files to those of \HPhi and supports many lattice models supported by \HPhi too;
\begin{itemize}
\item Hubbard and Heisenberg models,
\item Multi-band extensions of the Hubbard model,
\item Kondo lattice models.
\end{itemize}
Once the ground state is obtained, mVMC calculates the energy and the one-body and two-body Green's functions.

In 2016, PASUMS supported mVMC in improving the user interface and made it easy for a wide range of researchers; from the specialist (theoretical researcher) to the learner (undergraduate student).
mVMC gives the accurate ground state of larger systems than the ED method without sign problem, and hence it is one of the strongest tools to investigate the strongly correlated systems and the frustrated spin systems.

\subsubsection{TeNeS (2019, 2023)}
\begin{figure}
\includegraphics[width=\linewidth]{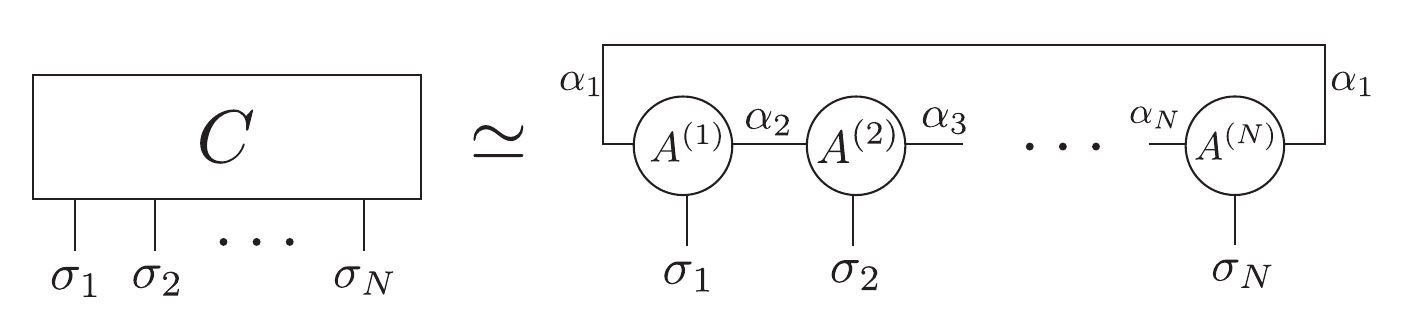}
\caption{Schematic picture of the MPS. The coefficient of a wave function $C$ is decomposed into a tensor network $A$.}
\label{fig:TN}
\end{figure}

Tensor networks (TNs) are another representation of the variational wavefunctions~\cite{Ors2014, Ors2019}.
Once a basis of the Hilbert space is given, for example, direct product state of up spin and down spin, the wave function is expanded as
\begin{equation}
\Ket{\Psi} = \sum_{\{\sigma_i\} = \uparrow,\downarrow} C_{\sigma_1\sigma_2\dots\sigma_N}\Ket{\sigma_1\sigma_2\dots\sigma_N}.
\end{equation}
The coefficient $C$ is an exponentially large tensor with $N$ indices and $d^N$ elements ($d=2$ is the degree of freedom on each site), and the TN framework represents this as a network of many small tensors.
For example, a wave function of a spin chain with $N$ sites under the periodic boundary condition is well represented as a product of $N$ small tensors as
\begin{equation}
C_{\sigma_1 \sigma_2 \dots \sigma_N}
\simeq
\sum_{\{\alpha_i\}}
A^{(1)\sigma_1}_{\alpha_1\alpha_2} A^{(2)\sigma_2}_{\alpha_2 \alpha_3} \dots A^{(N)\sigma_N}_{\alpha_N\alpha_1}.
\end{equation}
Fixing the values of $\set{\sigma}$ reduces the small tensors to matrices, and their product gives the element of $C$.
Therefore, this is called a matrix product state (MPS) (Figure \ref{fig:TN}).
Each tensor has three indices; one index (called a physical index) represents the local degree of freedom and two (called a virtual index) connect tensors to another.
The dimension of virtual indices $D$ is called the bond dimension and controls the accuracy of the TN state.
The TN can exponentially reduce the number of elements from $d^N$ to $dD^2N$.
Additionally, by imposing the translational symmetry of the state and introducing the sub-lattice order, the number of independent tensors can be further reduced.
For example, when all the tensors are common, the coefficient is represented as
\begin{equation}
C_{\sigma_1 \sigma_2 \dots \sigma_N}
\simeq
\sum_{\{\alpha_i\}}
A^{\sigma_1}_{\alpha_1 \alpha_2} A^{\sigma_2}_{\alpha_2 \alpha_3} \dots A^{\sigma_N}_{\alpha_N \alpha_1}.
\end{equation}

As for the chain lattice, systems, tensor networks on the square lattice can represent wave functions on two-dimensional lattices.
It is called the tensor product state (TPS)~\cite{Nishino2001} or the pair entanglement product state (PEPS)~\cite{Verstraete2004}.
By repeating tensors, a wave function on an infinitely large lattice can be represented by TPS, which is called the infinite TPS (iTPS).

TeNeS~\cite{TeNeS} is software for obtaining the iTPS representing the ground state of quantum lattice models.
It was released with the help of PASUMS in 2019.
To optimize the tensors, TeNeS performs the imaginary-time evolution of the state as $\ket{\Psi} = \lim_{n\to\infty} \left(\exp(-\tau H)\right)^n \ket{\psi_0}$ with small time step $\tau$~\cite{Jiang2008, Jordan2008, Orus2009, Phien2015}.
Tensor contraction of iTPS, an infinitely large tensor network, is performed with the corner transfer matrix renormalization group (CTMRG) method~\cite{Nishino1996, Orus2009}.
While TeNeS operates on the square lattice only, it takes other lattices such as a triangular lattice
%by considering long-range interactions on a square lattice.
by regarding it as a square lattice with next-nearest neighbor (or even further neighbor) interactions.
% As for other PASUMS software, TeNeS also offers utility tools for preparing the input files for widely used models and lattices from one simple file.
TeNeS, like other PASUMS software, provides utility tools that help users to generate input files for calculating widely used models and lattices:
\begin{itemize}
   \item general spin XXZ model
   \item Bose-Hubbard model
\end{itemize}
on
\begin{itemize}
   \item hypercubic lattice
   \item triangular lattice
   \item honeycomb lattice
   \item kagome lattice.
\end{itemize}

In TeNeS, basic tensor operations such as singular value decomposition are implemented with mptensor~\cite{mptensor, mptensor_Github}, a tensor library supporting the OpenMP/MPI hybrid parallelization via LAPACK and ScaLAPACK.
Therefore, TeNeS supports parallel calculations even in massively parallel computers and can perform heavy calculations with large bond dimensions.

In 2023, PASUMS supported TeNeS in implementing more calculation modes; the real-time evolution calculation~\cite{Czarnik2019} and the finite temperature calculation~\cite{Kshetrimayum2019}.
The real-time evolution is achieved simply by replacing $\tau$ in the time evolution operator $\exp(-\tau H)$ with $it$.
In the finite temperature calculation, a tensor network represents the density matrix representing the mixed state for the inversed temperature $\beta$, $\rho(\beta) \propto e^{-\beta H} = e^{-\beta H /2} \rho(0) e^{-\beta H /2}$, instead of the wave function $\ket{\Psi}$.
These extensions makes TeNeS a more useful tool in comparing with experiments, which often probe non-equilibrium or finite-temperature states.

\subsection{Machine Learning Related}

In addition to the applications developed through the advancement project mentioned above, first-principles calculation and molecular dynamics packages widely used in the field of condensed matter, such as Quantum ESPRESSO, VASP, and LAMMPS, are pre-installed on the ISSP supercomputer. The ISSP supercomputer's job scheduler is equipped with features for bulk jobs and array jobs, allowing for the execution of multiple hybrid parallel applications and providing an environment for various exhaustive calculations, such as parameter scans. In PASUMS, the application software that utilizes such an environment is developed based on machine learning, optimization problems, and data-driven approaches.

\subsubsection{abICS (2019, 2022)}
\begin{figure}
\begin{center}
\includegraphics[width=0.8\textwidth]{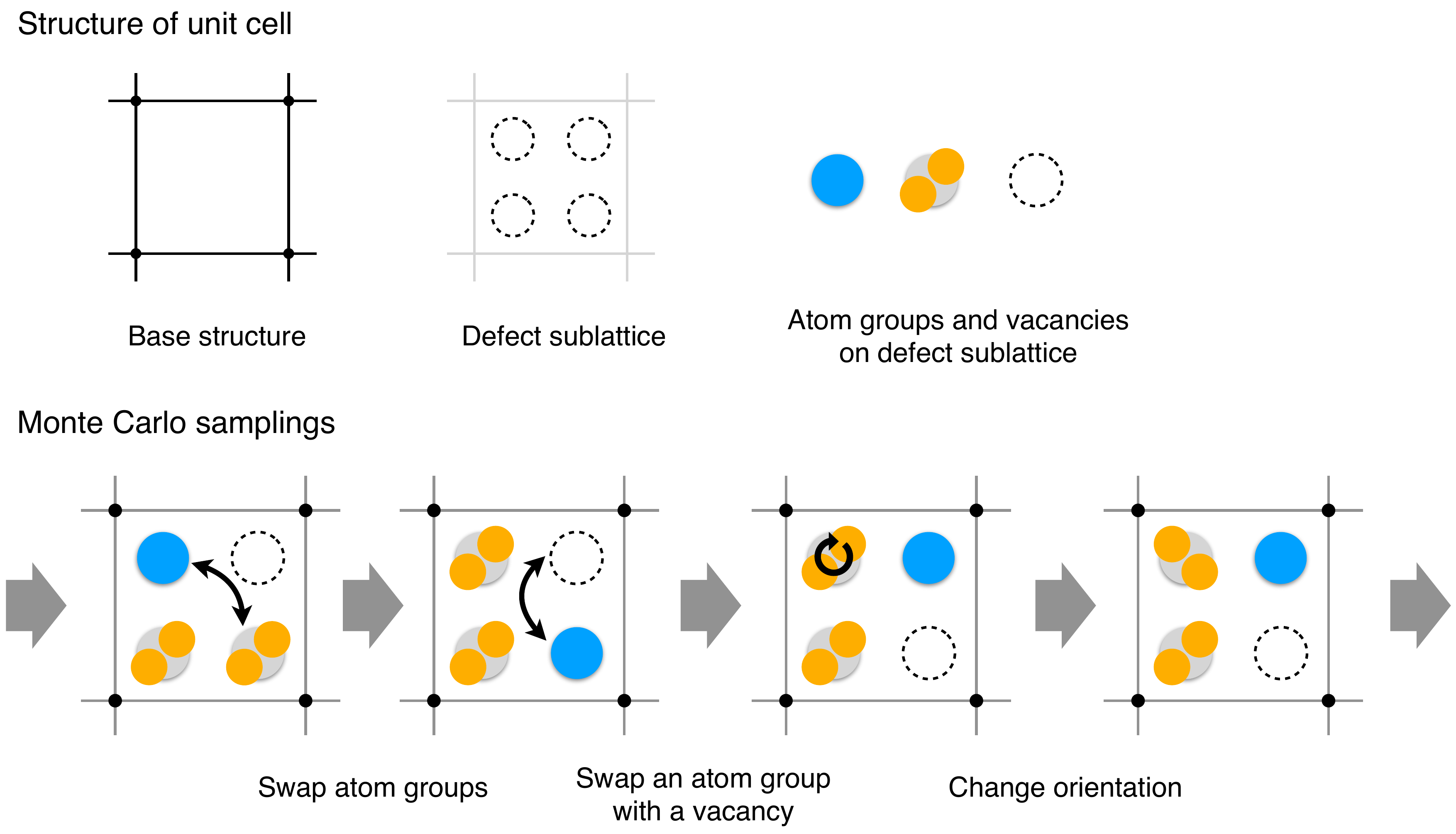}
\end{center}
\caption{Schematic figures of the structure of a unit cell and the sequence of Monte Carlo samplings. A unit cell is comprised of a base structure and a set of defect sublattices that accommodate atom groups and vacancies (upper figures). A sequence of configurations are generated according to Monte Carlo samplings that involve trial steps of exchanging atom groups and vacancies, and changing the orientation of atom groups (lower figure).}
\label{fig:abics}
\end{figure}

The configurational disorder in various functional materials is a crucial factor in determining material properties.
The ability to simulate such disorder is important for the prediction of properties, design of materials, and comparison with experiments.
Applying first-principles calculations directly to their evaluation generally results in enormous computational costs, so the importance sampling method is adopted.
When the disorder is not completely random but some short-range order exists, it is necessary first to perform thermodynamic sampling for configurations to clarify the structural order.
Traditionally, lightweight effective models have been derived by cluster expansion methods that fit the results of the first-principle calculations, to which the Monte Carlo sampling is applied.
However, for complex ionic crystals having many components and multiple sublattice structures, the cluster expansion becomes combinatorially demanding and intractable.

abICS~\cite{abics} is a framework to perform the direct statistical thermodynamic sampling by combining the high-throughput first-principles calculations, the parallel extended ensamble algorithms, and the on-lattice neural network models concurrently improved in an active learning settings (Figure~\ref{fig:abics}).
It makes efficient use of massively-parallel supercomputers to examine such multi-component multi-sublattice systems without phenomenological models~\cite{Kasamatsu2022,abics_stam_methods2023}.
In the fiscal year 2019, the original program for the thermodynamic sampling is extended to support the first principles calculation software including Quantum ESPRESSO and OpenMX in addition to VASP in a modular structure.
For the sampling algorithms, the replica exchange Monte Carlo method and the population annealing Monte Carlo method are implemented.
A new user interface is introduced for the ease of use: The overall procedure is controlled by the input file in TOML format, and the tools to generate input data are prepared.
In the fiscal year 2022, it is enhanced by the acceleration using the neural network model and the active learning.
The supported library includes aenet that is integrated through the file-I/O based interface as well as the aenet-LAMMPS Python module.
The grand canonical sampling is implemented that allows for changes in the composition.
abICS will be immensely useful for making efficient use of next-generation large-scale computers due to its multi-layered parallelism.
By employing the modular coding practice, it should be relatively easy to implement interfaces to other first-principles calculation software, or to introduce new sampling algorithms.

\subsubsection{2DMAT (2020, 2021, 2024)}
Formulation of a reliable and efficient method for the analysis of experimental data is one of the significant issues in scientific research.
In the analysis procedure, one often want to obtain the parameter $X$ characterizing the model from the experimentally observed quantities $D_\text{ex}$.
Generally, $X$ and $D$ are vectors and the dimension of $D$ is greater than that of $X$.
Typically, it can be regarded as an {\it inverse problem}.
Suppose that we can solve the {\it direct problem} of calculating the outcome of the measurement $D_\text{cal}(X)$ when the model parameter is $X$, and we can compute the loss function, $F(X)$, defined as some properly defined distance between the calculated outcome $D_\text{cal}(X)$ and the experimentally observed result $D_\text{ex}$.
Then, the inverse problem is to find the optimal parameter value $X^\ast$ that minimizes the loss function $F(X)$.

\begin{figure}
\begin{center}
\includegraphics[width=0.5\textwidth]{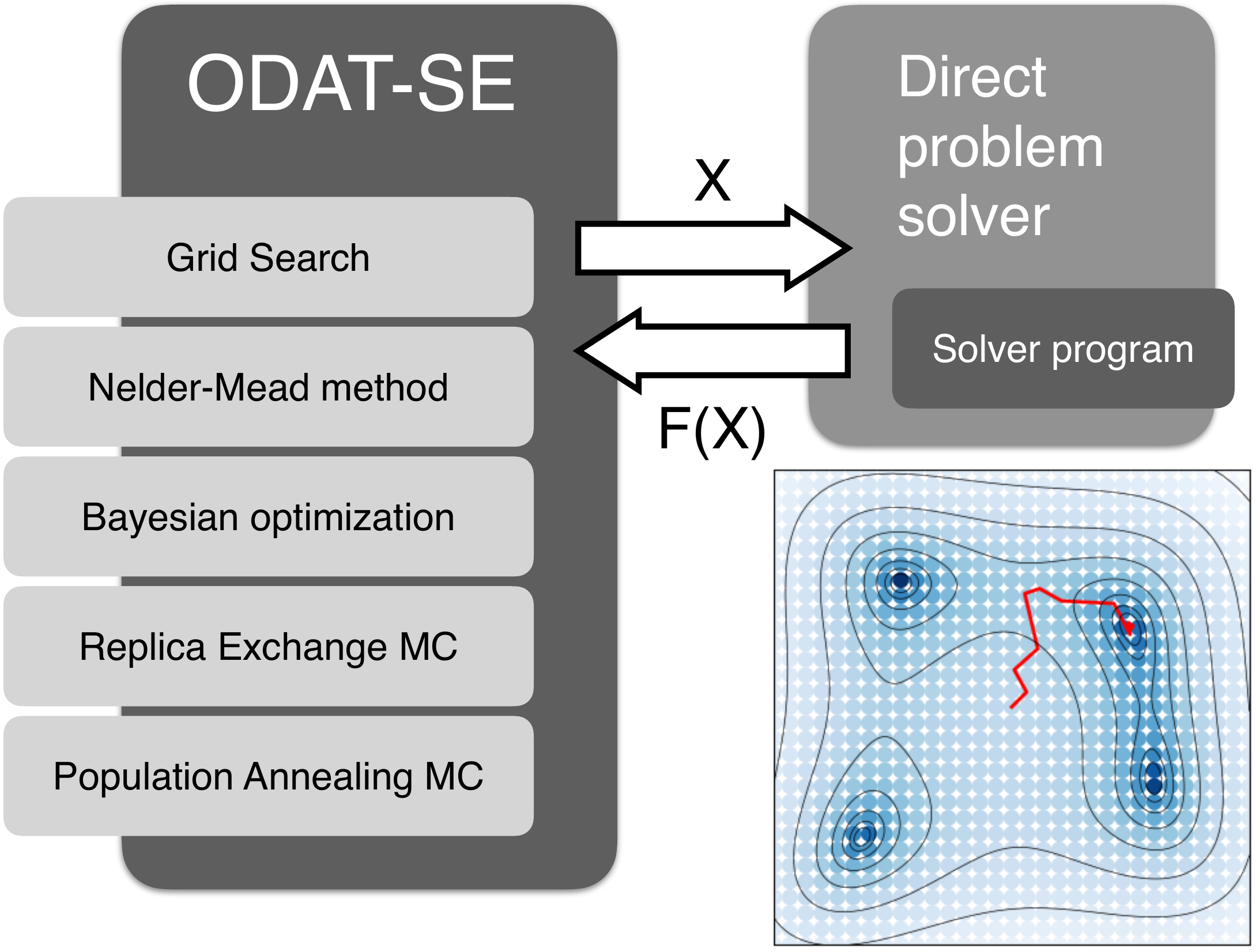}
\end{center}
\caption{Schematic view of ODAT-SE, an open framework for data analysis. For a given direct problem solver, ODAT-SE applies search algorithms to find optimal parameter values $X^*$ that minimize the loss function $F(X)$. The direct problem solvers and the algorithms are modularized so that the users can provide them for their own problems.}
\label{fig:2DMAT}
\end{figure}

2DMAT/ODAT-SE is an exploratory inverse problem analysis platform of the experimental data.
The implemented search algorithms include the grid-based search, the Nelder-Mead method, the Bayesian optimization using the PHYSBO library mentioned below, the replica exchange Monte Carlo method~\cite{Hukushima1996}, and the population annealing Monte Carlo method~\cite{Hukushima2003}.
By combining multiple methods, the global search of solutions can be attained, instead of being limited to local minima.
It can be applied to the analysis that takes account of the experimental uncertainty.
Because of its development history, 2DMAT/ODAT-SE comes with a relatively large collection of direct problem solvers specialized in diffraction experiments for the two-dimensional material structure analyses (and thus named as 2DMAT).
However, in the latest version, it has been reformulated to a general framework for optimization problems required for solving inverse problems, and thus renamed as Open Data Analysis Tool for Science and Engineering (ODAT-SE).
In the FY2020 PASUMS project, the analysis program mainly for the TRHEPD experiment using the Nelder-Mead method and the grid search method is reorganized and extended to support Bayesian optimization and the replica exchange Monte Carlo method, and released as 2DMAT version 1.
In the FY2021 project, it is further extended to support more diffraction experiments including SXRD and LEED.
The population annealing Monte Carlo method is also added for the inverse problem solver algorithm that is suitable for large-scale parallel computations.
In the FY2024 project, it is reorganized as an open platform for data analysis by modularizing direct problem solvers and search algorithms, and released as ODAT-SE version 3.
Users can extend the definitions of direct problems and search algorithms, making it a versatile platform for inverse problem analysis. (Figure~\ref{fig:2DMAT}.)

\subsubsection{PHYSBO (2020)}
PHYSBO (optimization tools for PHYSics based on Bayesian Optimization)\cite{PHYSBO-paper} is a Python-based software tool designed for black-box optimization tasks specific to condensed matter physics, leveraging Bayesian optimization (BO).
BO\cite{garnett_bayesoptbook_2023} is a machine-learning-driven optimization method particularly suited to scenarios in physics, chemistry, and materials science where the target function (e.g., material properties) is complex, expensive to evaluate, or lacks an analytical expression. For example, in materials development, discovering optimal materials through trial-and-error can be formulated as a black-box optimization problem, with inputs such as composition, structure, and processing conditions, and outputs representing the desired material properties.

PHYSBO, initially developed as Python 3 software under a fiscal year 2020 project, extends the functionality of COMBO~\cite{combo-paper} to address specific needs in condensed matter physics effectively. 
Key advancements in PHYSBO include:
\begin{itemize} 
\item Implementation of MPI-based parallelization for acquisition function optimization, enabling massive scalability on supercomputing platforms such as the ISSP supercomputer.
\item Introduction of multi-objective optimization functionality, expanding its applicability to problems requiring the simultaneous optimization of multiple objectives. 
\item Development of a detailed user manual to facilitate adoption by the research community. 
\end{itemize}
These improvements address the computational bottlenecks of BO, making PHYSBO a highly efficient and versatile tool for complex optimization tasks.

In the field of physics, BO has already been applied to several problems, including autonomous X-ray scattering experiments, inverse scattering, crystal structure prediction, and effective model estimation. The PHYSBO package builds on these successes, further accelerating such studies and enabling the exploration of even more complex physical systems by leveraging supercomputers.

\subsection{Constructing environments}
\subsubsection{MateriApps Installer (2020)}
In materials science, numerical simulation has become indispensable for theoretical research. The advancement of computational materials science relies heavily on developing efficient algorithms for solving equations that describe material properties. Over the years, many excellent applications leveraging state-of-the-art algorithms have been developed. However, the accessibility of these tools to a broader audience, including experimentalists and corporate researchers, remains a challenge.

To address this, MateriApps~\cite{MateriApps,Konishi15}, a portal site for materials science simulations, was launched in 2013. 
MateriApps is a platform for disseminating information about computational materials science software to researchers. Despite this effort, one major obstacle for new users is installing and configuring these applications.
To mitigate this challenge, MateriApps LIVE!~\cite{MateriAppsLive,MAL-paper}, an environment that allows users to try out computational materials science applications on their devices quickly, was developed. MateriApps LIVE! is distributed as a Virtual Hard Disk Image (OVA) for VirtualBox and includes pre-installed applications, an operating system (Debian GNU/Linux), editors, visualization tools, and other essential environments. This setup lets users quickly establish a working computational environment, which benefits software training sessions and classroom settings.
However, while MateriApps LIVE! is well-suited for introductory purposes, it is limited in computational power since it operates as a virtual machine. To support users interested in conducting large-scale simulations, MateriApps Installer was developed~\cite{MateriAppsInstaller, MAL-paper} in 2013.

As part of the FY2020 Project for PASUMS, several significant updates to MateriApps Installer were performed: \begin{itemize} \item Organized the directory structure and scripts for better usability, \item Added comprehensive documentation and tutorials, \item Upgraded the list of supported software, \item Extended support for new hardware, including the ISSP supercomputer system B (ohtaka), \item Supported new compilers, such as GCC 10 and Intel oneAPI. \end{itemize}
MateriApps Installer includes installation scripts for a wide range of materials science applications, such as
\begin{itemize} 
\item Simulation tools: ALPS, ALPSCore, DSQSS, Quantum ESPRESSO, \HPhi, \Komega, LAMMPS, mVMC, OpenMX, RESPACK, and TeNeS, 
\item Libraries and tools: Boost, CMake, Eigen3, FFTW, GCC, Git, GSL, HDF5, LAPACK, libffi, OpenBLAS, OpenMPI, OpenSSL, Python3, ScaLAPACK, Tcl/Tk, and zlib. 
\end{itemize}
MateriApps Installer enables these materials science applications to be easily installed on the ISSP supercomputers, providing users with a ready-to-use environment for large-scale simulations. 

\subsubsection{HTP-Tools (2023)}
In recent years, approaches that leverage machine learning to predict physical properties and design new materials, collectively known as materials informatics, have gained significant popularity. A critical factor for achieving high accuracy in predictions and designs is the availability of large amounts of supervised data. From this perspective, databases such as Materials Project, which store crystal structures, experimental measurements, and first-principles calculation results, have been developed and are widely used.
However, many machine learning applications require particular materials data and physical property information, often unavailable in existing databases. Efficient methods and environments for generating such targeted training data would significantly advance materials informatics by providing a robust research foundation and enabling rapid progress in the field.

To meet this requirement, PASUMS has supported the development of high-throughput (HTP) tools designed for exhaustive data generation from crystal structures using first-principles calculations. One tool, cif2x, generates input files compatible with first-principles calculation software such as VASP, Quantum ESPRESSO, OpenMX, and AkaiKKR. Sample files and comprehensive tutorials are provided alongside cif2x to demonstrate its integration and practical application with these software packages.
Additionally, the project has introduced moller, a tool designed to automate batch job script generation, enabling large-scale computations on supercomputers. Though moller was developed independently from cif2x, it supports various computational solvers and is broadly applicable for general-purpose bulk calculations. Practical examples and tutorials illustrate its use with software such as HPhi and DSQSS, demonstrating how researchers can efficiently manage batch processing of multiple computational scenarios. Both cif2x and moller are distributed as open-source software under the GNU General Public License (GPL) version 3.0 and are pre-installed on systems such as the ISSP supercomputer.

As examples of their potential applications, these tools could be utilized to develop comprehensive computational materials science databases. Such databases would significantly accelerate materials informatics research and provide valuable resources to the broader scientific community. Additionally, extending moller's compatibility to supercomputing infrastructures beyond ISSP, such as those within the High-Performance Computing Infrastructure (HPCI), could further standardize large-scale computations. The availability of these tools thus supports researchers in generating diverse and extensive materials datasets, thereby substantially contributing to the advancement of materials informatics and computational materials science.

\section{Summary}
In this paper, we introduced the Project for Advancement of Software Usability in Materials Science (PASUMS), which ISSP carries out.
In the nationwide joint use of ISSP computer systems that began with the sharing of the hardware, the role of software and data is increasingly more essential, which is equally true for many other high-performance supercomputer systems worldwide.
The software developed or required by materials science researchers is highly diverse, making it challenging to cover comprehensively. However, about 10 years after the start of the PASUMS project, the open-source software born from this initiative has come to cover a significant portion of computational materials science. Recently, there has been a focus on enhancing the interoperability between different software. One of the future goals will be to build a framework for materials exploration using such ``integrated software'' and data repositories.

Additionally, various other initiatives have been launched, including a data repository project~\cite{datarepo-portal} initiated with the collaboration of data science and materials science, a portal site called MateriApps~\cite{MateriApps} for promoting software use, MateriApps LIVE!~\cite{MateriAppsLive,MAL-paper} which packages execution environments to facilitate the easy use of software, and hands-on workshops and lectures to promote software utilization. Those interested should refer to the references.

\section{Acknowledgement}
This paper introduced many open-source software projects. Needless to say, these owe much to the contributions of the original developers, project proposers, and collaborators. In many cases, ISSP has assisted in enhancing functionality and usability through manual preparation and other support efforts. We would like to express our gratitude to these individuals.

\bibliographystyle{tfnlm}
\bibliography{main}

\begingroup
\renewcommand{\arraystretch}{1.2}
\begin{table*}[t]
  \centering
  \begin{tabular}{|p{.2\textwidth}|p{.75\textwidth}|}
    \hline
    \multicolumn{2}{|l|}{
      First-Principles Calculation Related
    } \\ \hline
    abICS~\cite{abics} &
    Software framework for performing configuration sampling in disordered systems \\ \hline
    OpenMX~\cite{OpenMX} &
    First-principles calculation program using localized basis sets \\ \hline 
    RESPACK~\cite{RESPACK-paper, RESPACK} &
    First-principles calculation software for evaluating interaction parameters of materials \\ \hline
    ESM-RISM~\cite{ESMRISM} &
    Software describing the electronic states of electrodes and reactants using density functional theory and analyzing the distribution of electrolytes using classical solution theory \\ \hline
    \hline
    \multicolumn{2}{|l|}{
      Quantum Lattice Model Solver Related
    } \\ \hline
    H-wave~\cite{Hwave} &
    Calculation program using mean-field and random-phase approximation \\ \hline
    \HPhi~\cite{HPhi-paper,HPhi} &
    Program for finite temperature calculations using exact diagonalization and thermodynamic pure quantum states \\ \hline
    mVMC~\cite{mVMC-paper,mVMC} &
    Calculation program using the multi-variable variational Monte Carlo method \\ \hline
    DSQSS~\cite{DSQSS-paper,DSQSS} &
    Calculation program using the quantum Monte Carlo method \\ \hline
    DCore~\cite{DCore-paper,DCore} &
    Tool for calculations using the dynamical mean-field theory \\ \hline
    TeNeS~\cite{TeNeS-paper,TeNeS} &
    Calculation program using the tensor network method \\ \hline
    \hline
    \multicolumn{2}{|l|}{
      Others
    } \\ \hline
    \Komega~\cite{Komega-paper,Komega} &
    Solver library based on the Shifted-Krylov subspace method \\ \hline
    PHYSBO~\cite{PHYSBO-paper, PHYSBO} &
    O(N) Bayesian optimization library \\ \hline
    2DMAT~\cite{2dmat-paper, 2dmat} &
    Optimization and sampling framework using various methods \\ \hline
    MateriApps Installer~\cite{MAL-paper,MateriAppsInstaller} &
    Collection of shell scripts for installing open-source computational materials science applications and tools \\ \hline
    HTP-tools~\cite{htp_tools} &
    Tool scripts for exhaustive calculations (moller), support tools for creating input files for first-principles calculation software (cif2x). \\ \hline
  \end{tabular}
  \caption{List of software enhanced through the Software Development and Advancement Project (PASUMS).}
  \label{table-soft-list}
\end{table*}
\renewcommand{\arraystretch}{1.0}
\endgroup

\end{document}